# Design of nonuniformly spaced phase-stepped algorithms using their frequency transfer function

MANUEL SERVIN,* MOISES PADILLA, GUILLERMO GARNICA, AND GONZALO PAEZ

*Centro de Investigaciones en Optica A. C., Loma del Bosque 115, Col. Lomas del Campestre, Leon Guanajuato, Mexico, C. P. 37150.*
*Corresponding author: mservin@cio.mx*

**Here we show how to design phase-shifting algorithms (PSAs) for nonuniform phase-shifted fringe patterns using their frequency transfer function (FTF). Assuming that the nonuniform/nonlinear (NL) phase-steps are known, we introduce the desired zeroes in the FTF to obtain the specific NL-PSA formula. The advantage of designing NL-PSAs based on their FTF is that one can reject many distorting harmonics of the fringes. We can also estimate the signal-to-noise ratio (SNR) for interferograms corrupted by additive white Gaussian noise (AWGN). Finally, for non-distorted noiseless fringes, the proposed NL-PSA retrieves the modulating phase error-free, just as standard/linear PSAs do.**

Since the seminal work of Greivenkamp, "Generalized data reduction for heterodyne interferometry," in 1984 (G-PSA) [1], there has been an increasing interest for nonlinear phase-step PSAs (NL-PSAs) [2-10]. That is because (except for fringe projection profilometry) one cannot always guarantee pure-linear phase-shifted fringe-data [11]. Nowadays the best known NL-PSAs are: a) the generalized PSA (G-PSA) [1]; b) the advanced iterative algorithm (AIA) [2-5]; c) NL-PSAs which consider small phase-shift nonlinearity, approximated by a Taylor series [6-8], and d) the principal component analysis (PCA) of phase-shifted fringes [9,10]. The PCA-based PSA (PCA-PSA) normally requires many phase-shifted interferograms having at least one spatial fringe to obtain the estimated phase with small, non-zero error, even for noiseless fringes [9,10]. The PCA and the AIA were combined to obtain a faster AIA providing as initial condition the small-error estimation of the PCA-PSA [10]. Also we know that G-PSA, AIA, and PCA-PSA do not reject high-order harmonics of the fringes by design [1-10]. Moreover the G-PSA, AIA, and PCA-PSA do not give any estimation of the SNR of the demodulated signal [1-10]. People working on variations of G-PSA, AIA, and PCA-PSA can only make noise/harmonics robustness comparisons from specific simulated or experimental fringes [1-10]. In other words, using the available NL-PSA theory [1-10], one cannot know the SNR from basic stochastic process theory [12].

**Our contribution.** Here we present explicitly *N*-step NL-PSA formulas with a desired FTF spectral response. The NL-PSA's FTF reject the highest number of fringe harmonics for a given number of phase steps. For noiseless, non-distorted data, our NL-PSA formulas give the exact modulating phase, not just a good approximation (as other NL-PSAs do [6-10]). In simple terms, the phase recovered with our NL-PSA equals the error-free phase obtained by standard/linear PSAs for noiseless fringes [11]. Moreover, using the designed FTF, one can easily estimate the SNR from basic stochastic process theory [11,12]. This contrasts with G-PSA, AIA, and PCA-PSA which do not give any SNR figure-of-merit [1-10]. For amplitude-distorted fringes, our NL-PSA give much better results than G-PSA, AIA, or PCA-PSA because our NL-PSA explicitly rejects many harmonics. The only constrain to our FTF-based NL-PSA design is that one needs a previous estimate of the nonlinear phase-steps. But this is not difficult for spatial linear-carrier fringes, for which the Fourier method can be used, or by the use of the Carré nonlinear phase-step formula for temporal fringes with no spatial carrier [11].

**Linear phase-step PSAs.** Before going to the main contribution of this work, we briefly review the concept of the FTF for spectral analysis of linear phase-step PSAs. Let us start by the standard mathematical form for continuous phase-shifted fringes,

$$I(t;\varphi_{x,y}) = a_{x,y} + b_{x,y} \cos\left(\varphi_{x,y} + \omega_0 t\right). \quad (1)$$

The measuring phase is $\varphi_{x,y}$; the background of the fringes is $a_{x,y}$; the contrast is $b_{x,y}$, and the angular frequency is $\omega_0$. For notation economy, we will use ($\varphi$, $a$, $b$) instead of ($\varphi_{x,y}$, $a_{x,y}$, $b_{x,y}$). If our interferograms have amplitude distortion, then one must include the fringe harmonics as,

$$I(t;\varphi) = a + \sum_{k=1}^{\infty} b_k \cos\left[k\left(\varphi + \omega_0 t\right)\right]. \quad (2)$$

It is usual to express the *n*th sample $I(n;\varphi)$ as,

$$I(n;\varphi) = \int_{-\infty}^{\infty} I(t;\varphi)\delta(t-n)dt; \quad n \in \{0,1,...,N-1\}. \quad (3)$$

Being $\delta(t-n)$ the sampling Dirac delta function. Then a linear phase-stepped PSA may be written as,

$$Ae^{i\hat{\varphi}} = \sum_{n=0}^{N-1} c_n^* I(n;\varphi); \quad (c_n \in \mathbb{C}); i = \sqrt{-1}. \quad (4)$$

The asterisk denotes the complex conjugated. This system has the following impulse response,

$$h(t) = \sum_{n=0}^{N-1} c_n \delta(t-n). \quad (5)$$

Taking the Fourier transform of $h(t)$ one obtains the spectral response of the PSA (the FTF) as,

$$H(\omega) = F\{h(t)\} = \sum_{n=0}^{N-1} c_n e^{-in\omega}. \quad (6)$$

If the fringe data is corrupted by AWGN, the SNR-gain ($G_{SNR}$) for a linear $N$-step PSAs is given by [11],

$$G_{SNR} = \frac{\left|\sum_{n=0}^{N-1} c_n e^{-i\omega_0 n}\right|^2}{\sum_{n=0}^{N-1} |c_n|^2} \leq N. \quad (7)$$

The $G_{SNR}$ numerator is proportional to the energy of the demodulated signal at $\omega=\omega_0$, while the denominator is proportional to the filtered noise energy. The highest $G_{SNR}=N$ is obtained only for LS-PSA in which $\omega_0=2\pi/N$; otherwise $G_{SNR}<N$ [11]. For example, the 7-step linear least-squares PSA (LS-PSA) has the following FTF [11],

$$H(\omega) = \sum_{n=0}^{6} e^{i\omega_0 n} e^{-in\omega}; \quad (\omega_0 = 2\pi/7). \quad (8)$$

And the plot of the periodic $|H(\omega)|$ is shown in Fig. 1.

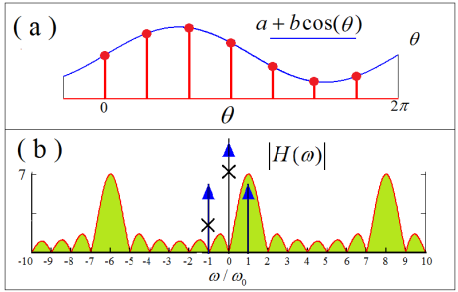

Fig. 1. Panel (a) shows 7 linear samples of a continuous fringe. Panel (b) shows the FTF of the 7-step linear LS-PSA [10]. Due to symmetry, this LS-PSA rejects many distorting harmonics at {-10,-9,-8,-7,-5,-4,-3,-2, 2, 3, 4, 5, 6, 7, 9, 10} within the frequency range [-10,10].

**Nonlinear phase-steps PSA**. We describe nonuniform temporal samples from Eq. (1) as,

$$I(t_n;\varphi) = \int_{-\infty}^{\infty} [a + b\cos(\varphi + \omega_0 t)]\delta(t-t_n)dt. \quad (9)$$

Being $t_n$ nonuniform sampling times. It is common practice to label the fringe samples by their nonlinear phase-steps as,

$$I(\theta_n;\varphi) = a + b\cos(\varphi + \theta_n); \quad (\theta_n = \omega_0 t_n). \quad (10)$$

Note that the angular frequency and sampling times are irrelevant. Therefore, from now on we will work with normalized frequency $\omega_0=1.0$ (radians/second). We then use $\theta_n$ instead of ($\theta_n/1.0$). We remark that $\theta_n$ are known. Figure 2 shows a possible realization of 9 nonuniform sampled fringe (red dots), and the Fourier spectra of the continuous-time fringe.

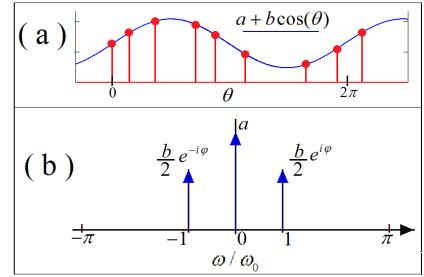

Fig. 2. Panel (a) shows 9 nonuniform phase-shifted samples of a continuous fringe at pixel ($x,y$). Panel (b) shows the Fourier spectra of the continuous fringe for $\theta \in \mathbb{R}$.

**FTF for nonuniform phase-stepped PSAs**. Our specific goal is to find a NL-PSA as a linear combination of the nonuniform phase-stepped interferograms as,

$$Ae^{i\hat{\varphi}} = \sum_{n=0}^{N-1} c_n^* I(\theta_n;\varphi); \quad (c_n \in \mathbb{C}). \quad (11)$$

Following the same receipt as linear PSAs [11], the NL-PSA's impulse response is given by,

$$h(t) = \sum_{n=0}^{N-1} c_n \delta(t-\theta_n/1.0). \quad (12)$$

And its FTF is given by,

$$H(\omega) = F\{h(t)\} = \sum_{n=0}^{N-1} c_n e^{-i\theta_n \omega}. \quad (13)$$

If the fringe data is corrupted by AWGN, the SNR-gain ($G_{SNR}$) for a $N$-step NL-PSAs is given by,

$$G_{SNR} = \frac{\left|\sum_{n=0}^{N-1} c_n e^{-i\theta_n}\right|^2}{\sum_{n=0}^{N-1} |c_n|^2} \leq N. \quad (14)$$

The equality is obtained if, and only if, $\theta_n = 2\pi n/N$; reducing to the standard linear LS-PSA (see Eq. (7)).

**Three step NL-PSA**. The *minimum* (normalized) quadrature conditions are,

$$H(-1) = 0; \quad H(0) = 0; \quad H(1) = 1. \quad (15)$$

According to these $H(\omega)$ constraints, one needs to solve for ($c_0,c_1,c_2$), for the *known* phase-steps ($\theta_0 = 0, \theta_1, \theta_2$) as,

$$\sum_{n=0}^{2} c_n e^{i\theta_n} = 0; \quad \sum_{n=0}^{2} c_n = 0; \quad \sum_{n=0}^{2} c_n e^{-i\theta_n} = 1. \quad (16)$$

Which can be rewritten in matrix form as,

$$\begin{bmatrix} 1 & e^{i\theta_1} & e^{i\theta_2} \\ 1 & 1 & 1 \\ 1 & e^{-i\theta_1} & e^{-i\theta_2} \end{bmatrix} \begin{bmatrix} c_0 \\ c_1 \\ c_2 \end{bmatrix} = \begin{bmatrix} 0 \\ 0 \\ 1 \end{bmatrix}. \quad (17)$$

For $\theta_0 \neq \theta_1 \neq \theta_2$ this 3-by-3 matrix is nonsingular, and one finds $(c_0, c_1, c_2)^T$. The explicit NL-PSA formula is then,

$$Ae^{i\hat{\varphi}} = c_0^* I(\theta_0;\varphi) + c_1^* I(\theta_1;\varphi) + c_2^* I(\theta_2;\varphi). \quad (18)$$

This 3-step NL-PSA is error-free ($\hat{\varphi} = \varphi$) for noiseless, non-distorted, temporal fringes. Figure 3(a) shows 3 nonuniform phase samples, and the FTF which phase demodulate them.

**Five phase-steps NL-PSA**. With more than 3 nonuniform phase-stepped fringes, one may reject more fringe harmonics. For example if we want the FTF to have the following constraints,

$$H(-2) = 0; H(-1) = 0; H(0) = 0; H(1) = 1; H(2) = 0. \quad (19)$$

One would need 5 coefficients $(c_n)$ as,

$$\begin{bmatrix} 1 & e^{2i\theta_1} & e^{2i\theta_2} & e^{2i\theta_3} & e^{2i\theta_4} \\ 1 & e^{i\theta_1} & e^{i\theta_2} & e^{i\theta_3} & e^{i\theta_4} \\ 1 & 1 & 1 & 1 & 1 \\ 1 & e^{-i\theta_1} & e^{-i\theta_2} & e^{-i\theta_3} & e^{-i\theta_4} \\ 1 & e^{-2i\theta_1} & e^{-2i\theta_2} & e^{-2i\theta_3} & e^{-2i\theta_4} \end{bmatrix} \begin{bmatrix} c_0 \\ c_1 \\ c_2 \\ c_3 \\ c_4 \end{bmatrix} = \begin{bmatrix} 0 \\ 0 \\ 0 \\ 1 \\ 0 \end{bmatrix}. \quad (20)$$

One may also change the desired FTF's constraints to,

$$H(-1) = 0; H(0) = 0; H(1) = 1; H(2) = 0; H(3) = 0. \quad (21)$$

Then the five coefficients $(c_n)$ change to,

$$\begin{bmatrix} 1 & e^{i\theta_1} & e^{i\theta_2} & e^{i\theta_3} & e^{i\theta_4} \\ 1 & 1 & 1 & 1 & 1 \\ 1 & e^{-i\theta_1} & e^{-i\theta_2} & e^{-i\theta_3} & e^{-i\theta_4} \\ 1 & e^{-2i\theta_1} & e^{-2i\theta_2} & e^{-2i\theta_3} & e^{-2i\theta_4} \\ 1 & e^{-3i\theta_1} & e^{-3i\theta_2} & e^{-3i\theta_3} & e^{-3i\theta_4} \end{bmatrix} \begin{bmatrix} c_0 \\ c_1 \\ c_2 \\ c_3 \\ c_4 \end{bmatrix} = \begin{bmatrix} 0 \\ 0 \\ 1 \\ 0 \\ 0 \end{bmatrix}. \quad (22)$$

Figures 3(b) and 3(c) show 5 nonuniform samples of a phase-shifted fringe, and two different FTFs which phase-demodulate the 5 fringe-samples. The resulting 5-sample NL-PSA formula is given by,

$$Ae^{i\hat{\varphi}} = \sum_{n=0}^{4} c_n^* I(\theta_n;\varphi). \quad (23)$$

These two 5-step NL-PSAs are also error-free ($\hat{\varphi} = \varphi$) for noiseless, non-distorted, fringes.

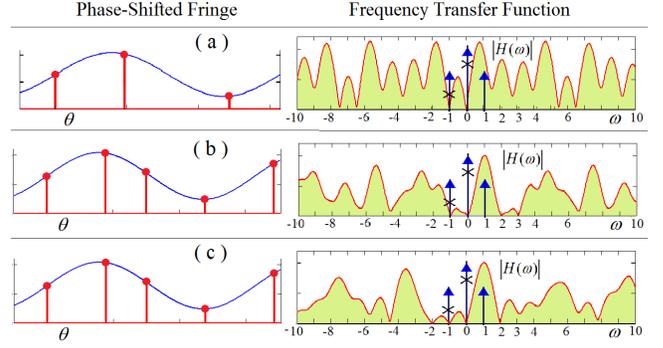

Fig. 3. As far as we know, this is the first time that FTFs for NL-PSAs are shown. Panel (a) shows 3 nonlinear phase-steps fringe data, and its corresponding FTF with zeroes at $\omega$={-1,0}. Panel (b) shows 5 nonlinear phase-steps and its FTF with zeroes at $\omega$={-1,0,2,3}. Panel (c) show the same data as Panel (b), but now the FTF is designed to have zeroes at $\omega$={-2,-1,0,2}.

**Harmonics rejection of the FTF-based NL-PSA**. Unlike uniformly sampled linear PSA, $H(\omega)$ is not periodic for NL-PSAs. This means that the only harmonics rejected by the NL-PSA are the ones explicitly rejected in the design of its FTF. For instance consider a 7-step FTF $H(\omega)$ having the following spectral zeroes:

$$\begin{aligned} H(-2) &= 0; \quad H(-1) = 0; \quad H(0) = 0; \quad H(1) = 1; \\ H(2) &= 0; \quad H(3) = 0; \quad H(4) = 0. \end{aligned} \quad (24)$$

As Fig. 4 shows, this is a symmetrical distribution of spectral zeroes around $\omega$=1. The nonlinear 7 phase-steps used are $(\theta_n)$ =(0, 0.78, 1.81, 3.11, 4.54, 5.93, 7.24). Note that the only fringe-harmonics rejected are the ones explicitly designed into its FTF.

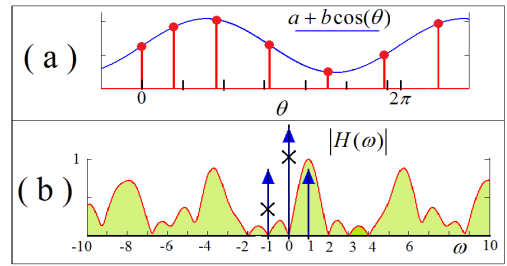

Fig. 4. Panel (a) shows seven nonlinear phase-steps of a fringe at pixel (x,y). The phase-shifting coordinate is $\theta$ (radians). Panel (b) shows the FTF of the NL-PSA which demodulate, *error-free*, the 7 temporal samples. Note that this FTF have zeroes at $\omega$={-2,-1, 0,+2,+3,+4}.

Both Fig. 1 and Fig. 4 correspond to 7-step PSAs. In Fig. 1 the LS-PSA's FTF has zeroes at: {-10, -9, -8, -7, -5, -4, -3, -2, -1, 0, 2, 3, 4, 5, 6, 7, 9, 10}. In contrast our NL-PSA has only the designed FTF's zeros at {-2, -1, 0, 2, 3, 4} in the same range. In other words the use of nonuniform phase-steps reduces the harmonic rejection capacity of the NL-PSA. Nevertheless, this is much better than G-PSA, AIA, and PCA-PSA which only reject the quadrature components of the fundamental fringe frequency [1-10]. This is a serious disadvantage for amplitude-distorted fringe-data because all distorting harmonics survive. Even worst the AIA and the PCA-PSA must pre-filter its background signal at $\omega=0$ [5,9,10].

**Computer simulation**. We simulated 7 fringe patterns with the aforementioned phase-steps $(\theta_n)=(0, 0.78, 1.81, 3.11, 4.54, 5.93, 7.24)$ and designing the FTF to have the zeroes in Eq. (24). The resulting FTF's complex coefficients are $(c_n)=\{-0.06, 0.21-0.21i, -0.05-0.2i, -0.22, -0.04-0.22i, 0.23+0.08i, -0.07-0.1i\}$. The resulting NL-PSA is then,

$$Ae^{i\hat{\varphi}} = \sum_{n=0}^{6} c_n^* I(\theta_n;\varphi) \ . \quad (24)$$

The first sample has the lowest weight $|c_0|=0.06$, meaning that its information is taken less into account. Two out of seven noiseless fringes, and its demodulated phase are shown in Fig. 5.

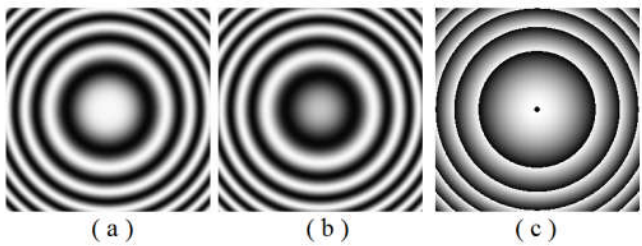

Fig. 5. Panels (a)-(b) show 2 out of 7 noiseless, nonlinear phase-stepped fringes. Panel (c) shows the *error-free* estimated phase.

We remark that the estimated phase for our FTF based NL-PSAs, is mathematically *error-free* ($\hat{\varphi}=\varphi$), whenever the phase-steps ($\theta_n$) among the interferograms are known accurately. However, error-free phase estimation is not mathematically guaranteed in PCA-PSA [10]. In the case of AIA and noiseless fringes, it normally takes many iterations to reach an error-free phase estimation [10]. In other words, for noiseless fringes, the estimated phase recovered by our NL-PSAs is as good as the one obtained by standard linear PSAs [11]. Of course, for fringes corrupted by AWGN and harmonics, we obtain a distorted demodulated phase, as it is the case for standard linear PSAs [11].

**SNR gain for our specific NL-PSA**. For our specific case with $(\theta_n)=(0, 0.78, 1.81, 3.11, 4.54, 5.93, 7.24)$, the SNR-gain is given by,

$$G_{SNR} = \frac{\left|\sum_{n=0}^{6} c_n e^{-i\theta_n}\right|^2}{\sum_{n=0}^{6} |c_n|^2} = 5.142. \quad (25)$$

Resulting in a 27% SNR-gain reduction with respect to a 7 samples linear LS-PSA ($G_{SNR}=7$).

**Comparison against PCA-PSA**. Before concluding we show a comparison of our FTF based NL-PSA and the PCA-PSA. It is well known that the PCA-PSA does not give, in general, an error-free phase estimation, even for noiseless nonlinear phase-stepped fringes [9,10]; this can be seen in Fig. 6. However the approximate PCA-PSA's solution may be used as initial condition for the AIA. Then the AIA, after several iterations, converges to an almost error-free phase estimation, for noiseless fringes [10].

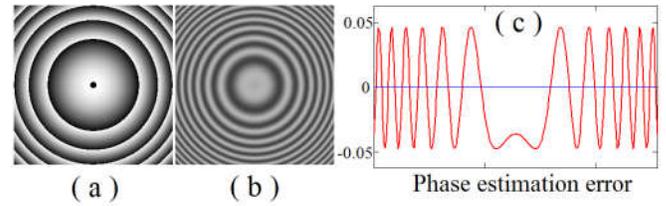

Fig. 6. Panel (a) shows the PCA-PSA's estimated phase from the noiseless fringe-data in Fig. 5. Panel (b) shows the phase estimation error. Panel (c) shows a central cut of the phase estimation error in radians. The phase error of our NL-PSA is about $10^{-15}$ (a numerical zero).

**Conclusions**. The herein proposed NL-PSA theory is a key contribution to nonlinear phase-steps interferometry in the sense that:

1) As far as we know, the spectral response (the FTF) for NL-PSAs was obtained for the first time. This FTF is in turn used to find the corresponding *N*-step NL-PSA formula.

2) For a given number of nonlinear phase-steps ($\theta_n$), our NL-PSA has the highest fringe harmonics rejection. In contrast, the G-PSA, AIA or PCA-PSA do not reject, by design, higher order harmonics of the fringe data [1-10].

3) Our FTF-based NL-PSA give us as a bonus, the signal-to-noise ratio gain ($G_{SNR}$) for fringes corrupted by AWGN. This contrast with G-PSA, AIA, and PCA-PSA which *do not give* a SNR estimate [1-10] from basic stochastic process theory [11,12].

4) Finally, our FTF-based NL-PSA design recovers the demodulated phase *error-free* for noiseless, non-distorted fringes. This not being the case for PCA-PSA [9,10].

**Full References**